\documentclass[
    useAMS,
    usenatbib,
    referee
]{biom}

\usepackage{graphicx}

\usepackage{amsmath}
\usepackage{amssymb}
\usepackage{enumerate}

\usepackage{tabularx}
\usepackage{longtable}
\usepackage{ltablex}
\keepXColumns

\usepackage{multirow}
\usepackage{threeparttable}
\usepackage{booktabs}
\usepackage{xr}

\usepackage[scr=rsfs]{mathalpha}

\usepackage{comment}
\usepackage{url}

\usepackage{soul}
\sethlcolor{yellow}

\usepackage{xcolor}

\theoremstyle{plain}
\newtheorem{proposition}{Proposition}

\begin{document}
\baselineskip=16pt
\begin{center}
	{\Large \bf Probability of Success under Population Shift for Phase III Trial Planning} 
\end{center}

\begin{center}
	Xinyi He$^{1,2}$, Jianchang Lin$^{3}$, Rachael Liu$^{3}$,  Ying Yuan$^{1, *}$,\\
\end{center}

\noindent$^1$Department of Biostatistics, The University of Texas MD Anderson Cancer Center, Houston, TX 77030, USA\\
$^2$Graduate School of Biomedical Sciences, The University of Texas Health Science Center at Houston, Houston, TX 77030, USA \\
$^3$ Bristol Myers Squibb, East Cambridge, MA 02141, USA\\
$^*$ Email: yyuan@mdanderson.org

\baselineskip=24pt

\begin{center}
\textbf{Abstract}
\end{center}
Probability of success (PoS) is widely used to inform go/no-go decisions for phase III trials based on phase II evidence. Conventional PoS calculations, however, implicitly assume that baseline covariate distributions are identical across phases. When covariate shift exists between phase II and phase III populations, unadjusted PoS may be biased and lead to miscalibrated decision-making. We propose a Bayesian  framework for PoS evaluation that explicitly accounts for covariate shift between phase II and III population prior to phase III initiation. The method adopts a transportability perspective, adjusting phase II evidence to reflect an anticipated phase III target population using available target covariate information. The framework accommodates both individual-level and aggregate-level phase III covariate data. Simulation studies demonstrate that, compared with the conventional approach, the proposed method improves PoS accuracy and decision calibration when covariate shift is present, while maintaining comparable performance when populations are aligned. We illustrate the proposed methodology using clinical trial data from systemic lupus erythematosus.

\vspace{5mm}
\clearpage

\section{Introduction}
Drug development is heavily burdened by prohibitively high development costs and substantial  failure rates. The average cost of a confirmatory phase III trial ranges between \$11.5 million and \$52.9 million \citep{sertkaya2016key}. Despite such substantial investments, phase III failure rates remain as high as 50\% \citep{hwang2016failure}. 
An essential strategy to curb these costly late-stage failures is to establish more rigorous and reliable go/no-go criteria informed by phase II data. Specifically, if the predicted success rate of a phase III trial based on phase II evidence is low, the phase III trial should not be initiated.

The Probability of Success (PoS) has gained popularity in drug development as a quantitative framework for making go/no-go decisions regarding phase III trials \citep{o2005assurance}. 
PoS based on observd data $D$ can generally be defined as $$\operatorname{PoS}(D)=\int\Pr(\text{Success}\mid\Delta)p(\Delta\mid D)d\Delta,$$ where ``$\text{Success}$" indicates that the phase III trial has met prespecified criteria (e.g., reaching a certain p-value threshold \citep{chuang2006sample} or satisfying Bayesian success criteria \citep{ibrahim2015bayesian}), and $\Delta$ represents the true treatment effect. For the sake of simplicity, we assume here that $D$ comes from the preceding phase II trial, noting, however, that it can represent any prior information (e.g., phase I cohort expansion or real-world data). PoS can also be interpreted as the expected power under the prior $p(\Delta\mid D)$, a concept closely related to Bayesian power \citep{spiegelhalter2004bayesian}. 

A range of methodologies has been proposed to enhance the accuracy of PoS computation. \citet{ibrahim2015bayesian} proposed a fully Bayesian approach that allows for the inclusion of covariates and historical data via the power prior. \citet{proper2024predicting} augmented the control arm of a randomized phase II trial with external patient-level data to enhance the posterior distribution estimation for PoS calculation using a propensity-score-based meta-analytic predictive prior. 
More recently, \citet{sachdeva2025assessing} proposed a unified Bayesian approach to PoS that coherently combines same-endpoint or surrogate-endpoint data with information from multiple historical trials. Methods for calculating PoS using other information sources have also been investigated, including early-phase pharmacodynamics \citep{chuang2011quantitative} and expert opinion \citep{hampson2022improving}.

Furthermore, alternative definitions of ``Success" have been explored to expand the applicability of the PoS framework. \citet{saint2018decision} adopted a composite success criterion that requires not only statistical significance on the primary endpoint but also clinical relevance and a favorable benefit-risk balance in subsequent pivotal studies. \citet{hampson2022new} proposed a more comprehensive framework defining success by positive pivotal trial outcomes, regulatory approval, and the fulfillment of target product profile requirements for market access. 

A key challenge in applying PoS to inform go/no-go decisions for phase III trials is that population characteristics often shift between phase II and phase III cohorts \citep{kirby2012discounting}. For example, pivotal phase III study populations are typically more heterogeneous than phase II cohorts \citep{kirby2012discounting}, which may result in smaller treatment effect.  When this occurs, $p(\Delta\mid D)$ fails to reflect the true distribution of $\Delta$ in the phase III population, leading to miscalibrated decision-making for the pivotal trial. Several approaches have been proposed to address this challenge. For instance, \citet{kirby2012discounting} suggested discounting the phase II treatment effect through a multiplicative adjustment with a retention factor to compute PoS. \citet{ibrahim2015bayesian} recommended reducing the weight of the historical likelihood in the power prior for the covariate distribution when the future-trial population differs from the historical cohort; however, their approach does not explicitly specify how to address the resulting bias. \citet{gotte2017} accounted for PoS bias caused by biomarker subgroup selection in phase II. Additionally, \citet{sabin2014quantitative} provided a Bayesian predictive modeling workflow that fits a meta-regression to explicitly allow study-level covariates and prognostic characteristics to map phase II outcomes to anticipated phase III outcomes. 

In this study, based on the trial transportability framework, we propose the transported Probability of Success (tPoS), a method for calculating the PoS of pivotal phase III trials when baseline characteristics shift between phase II and phase III populations for binary and continuous endpoints. By transporting evidence from the phase II source population to the anticipated phase III target population via weighting or moment matching, we effectively account for covariate shift. The tPoS approach enables an accurate PoS evaluation tailored to the intended target population. Crucially, the proposed framework accommodates both individual-level data and aggregate-level covariate summaries for the phase III population.

The remainder of this paper is organized as follows. 
Section 2 introduces the formal definition of PoS and presents the proposed tPoS methodology. Section 3 evaluates the operating characteristics of tPoS using simulation.
In Section \ref{application}, we present an application of the tPoS methodology. We conclude with a discussion in Section \ref{discuss}.

\section{Method}
\subsection{{Probability of success}} \label{PoS intro}

Consider a randomized controlled phase~II trial with fixed sample size \(n_0\) and a subsequent randomized controlled pivotal phase~III trial with fixed sample size \(n_1\), evaluating the same treatments in non-overlapping participants. Let \(S\) denote the population-membership indicator, taking values
\(s\in\{0,1\}\), where \(s=0\) corresponds to the phase~II population and \(s=1\) to the phase~III population. For patient \(i=1,\ldots,n_s\) in phase \(s\), let \(z_{si}\in\{0,1\}\) denote treatment assignment (\(0=\) control, \(1=\) experimental), \(y_{si}\) the observed outcome, and \(\boldsymbol{x}_{si}=(x_{1,si},\ldots,x_{p,si})^\top\in\mathcal X\) a \(p\)-dimensional vector of baseline covariates including potential effect modifiers. Let \(
\boldsymbol y_s=(y_{s1},\ldots,y_{sn_s})^\top,~
\boldsymbol z_s=(z_{s1},\ldots,z_{sn_s})^\top
\). Let \(D_s=\{(y_{si},z_{si},\boldsymbol{x}_{si}):i=1,\ldots,n_s\}\) denote the data from phase \(s\), where \(D_0\) denotes the observed phase~II data and \(D_1\) the future phase~III data. Because treatment is assigned by randomization within each trial, \(z_{si}\) is independent of \(\boldsymbol{x}_{si}\) for \(s=0,1\). We denote the data without covariates by \(
D_s^{(y,z)}
=
\{(y_{si},z_{si}):i=1,\ldots,n_s\},
\)
$s=0,1$.

For binary endpoints, let
$\theta_z^{(s)}=\Pr(Y=1\mid Z=z,S=s)$
denote the marginal response probability.
For continuous endpoints, let
$\mu_z^{(s)}=E(Y\mid Z=z,S=s)$ and
$\{\sigma_z^{(s)}\}^2
=\operatorname{Var}(Y\mid Z=z,S=s)$
denote the marginal mean and variance, assumed finite.
Let $\Delta_s$ denote the marginal treatment effect
in phase $s$, defined as
$\Delta_s=\theta_1^{(s)}-\theta_0^{(s)}$
for binary endpoints and
$\Delta_s=\mu_1^{(s)}-\mu_0^{(s)}$
for continuous endpoints.
In particular, $\Delta_1$ is the treatment effect
in the phase~III population.
Without loss of generality, larger values of
$\Delta_s$ are taken to indicate greater efficacy
of the experimental treatment.
For the Bayesian analyses below, we use Bernoulli
and normal marginal outcome models for binary
and continuous endpoints, respectively.

Let \(\boldsymbol X\) denote a generic individual-level baseline
covariate vector, with realization \(\boldsymbol x\). The marginal
outcome distribution in phase \(s\) can be expressed as
\begin{align}
p_s(y\mid z)
&=
\int
p(y\mid z,\boldsymbol x,S=s)
p_s(\boldsymbol x)\,d\boldsymbol x,
\qquad s=0,1.
\label{eq:predictive}
\end{align}
Here, \(p_s(\boldsymbol x)=p(\boldsymbol x\mid S=s)\)
is the covariate distribution in phase \(s\). We assume that
the conditional outcome distribution is invariant across phases,
\(p(y\mid z,\boldsymbol x,S=0)
=p(y\mid z,\boldsymbol x,S=1)
=p(y\mid z,\boldsymbol x)\),
while \(p_0(\boldsymbol x)\) and \(p_1(\boldsymbol x)\)
may differ.

For the future phase~III trial, participants are independently
sampled from the target population, and treatment assignments
are generated independently of baseline covariates according to
the prespecified randomization scheme \(p(\boldsymbol z_1)\). For example, \(p(\boldsymbol z_1)=0.5\) for 1:1 randomization.

We define the oracle PoS as the probability
that the future phase~III trial satisfies a prespecified
success criterion under the true target outcome distribution:
\begin{align}
\operatorname{PoS}^{\mathrm{oracle}}
&=
\int\!\!\int
\mathbb I\left(
\mathrm{Success}\mid D_1^{(y,z)}
\right)
p_1(\boldsymbol y_1\mid\boldsymbol z_1)
p(\boldsymbol z_1)
\,d\boldsymbol y_1\,d\boldsymbol z_1,
\label{eq:oraclepos}
\end{align}
where
\(p_1(\boldsymbol y_1\mid\boldsymbol z_1)
=\prod_{i=1}^{n_1}p_1(y_{1i}\mid z_{1i})\)
is the outcome distribution under independent sampling
from the phase~III population, and
\(\mathbb I(\mathrm{Success}\mid D_1^{(y,z)})\)
indicates whether the future trial satisfies the prespecified
success criterion. In practice, $\operatorname{PoS}^{\mathrm{oracle}}$ is unknown because the true phase~III outcome distribution $p_1(\boldsymbol y_1\mid\boldsymbol z_1)$ is unknown. Moreover, at the time of PoS estimation, the phase~III trial has not yet been conducted, and only phase~II data are available. Thus, the key challenge is to estimate the phase~III outcome distribution $p_1(\boldsymbol y_1\mid\boldsymbol z_1)$ using phase~II data.


As noted previously, various criteria can be used to define ''Success''.
Here, we focus on the following Bayesian success criterion, as described in
recent FDA guidance on Bayesian methodology for clinical trials
\citep{FDA_Bayesian}: the experimental arm
is regarded as a success and superior to the control if $\Pr(\Delta_1>\delta_0\mid D_1^{(y,z)})>\gamma$, where $\delta_0$ is
a prespecified, clinically meaningful treatment effect (which is 0 for most
superiority trials), and $\gamma$ is a prespecified probability cutoff. Thus, in Equation~(\ref{eq:oraclepos}),
\begin{align}
\mathbb{I}(\text{Success}\mid D_1^{(y,z)})
=
\mathbb{I}\left\{
\Pr(\Delta_1>\delta_0\mid D_1^{(y,z)})>\gamma
\right\}.
\label{successdef}
\end{align}
Of note, the proposed methodology does not depend on a specific definition
of the success criterion. When appropriate, $p$-value-based success criteria
can also be used.

The standard approach to estimating $\operatorname{PoS}^{\mathrm{oracle}}$ assumes that $p_0(y\mid z)=p_1(y\mid z)$, allowing  $p_1(y\mid z)$, and hence $\operatorname{PoS}^{\mathrm{oracle}}$, to be estimated using phase II data. Specifically, this approach models the marginal outcome distribution in phase II using a parametric model $p(y\mid z,\boldsymbol\varphi)$ and constructs the following posterior predictive distribution:
\begin{align}
p^{\mathrm{Std}}
(\boldsymbol y_1\mid\boldsymbol z_1,D_0^{(y,z)})
&=
\int
p(\boldsymbol y_1\mid\boldsymbol z_1,\boldsymbol\varphi)
p(\boldsymbol\varphi\mid D_0^{(y,z)})
\,d\boldsymbol\varphi.
\label{eq:standardppd}
\end{align}
Under the assumption that $p_0(y\mid z)=p_1(y\mid z)$, the standard PoS is calculated by replacing the unknown phase III outcome distribution $p_1(\boldsymbol y_1\mid\boldsymbol z_1)$ with $p^{\mathrm{Std}}(\boldsymbol y_1\mid\boldsymbol z_1,D_0^{(y,z)})$, yielding
\begin{align}
\operatorname{PoS}^{\mathrm{Std}}(D_0^{(y,z)})
&=
\int\!\!\int
\mathbb I\left(
\mathrm{Success}\mid D_1^{(y,z)}
\right)
p^{\mathrm{Std}}
(\boldsymbol y_1\mid\boldsymbol z_1,D_0^{(y,z)})
p(\boldsymbol z_1)
\,d\boldsymbol y_1\,d\boldsymbol z_1.
\label{eq:standardppdos}
\end{align}

However, when covariate shift occurs between phase II and phase III trials (i.e., \(p_1(\boldsymbol x)\neq p_0(\boldsymbol x)\)), the marginal outcome distributions generally differ, $p_0(y\mid z)\neq p_1(y\mid z)$, as shown in equation~(\ref{eq:predictive}), even if the conditional outcome distribution remains invariant across phases with
$p(y\mid z,\boldsymbol x,S=0)=p(y\mid z,\boldsymbol x,S=1)=p(y\mid z,\boldsymbol x)$.
Consequently, the standard approach may yield a biased estimate of $\operatorname{PoS}^{\mathrm{oracle}}$. This limitation motivates our proposed tPoS, which accounts for covariate shift by adjusting phase II evidence to reflect the baseline covariate distribution of the phase III population.

\subsection{{Transported PoS (tPoS)}} \label{adjust}
\subsubsection{Assumptions and General Framework} \label{general}

We use the transportability framework to develop a Bayesian
predictive approach for estimating the target phase~III oracle
PoS defined in Equation~\eqref{eq:oraclepos}. Specifically,
we incorporate covariate-adjustment weights into the analysis
of phase~II outcomes to account for differences in baseline
covariate distributions between the source and target populations. 

We first establish the transportability identification
framework and then construct the covariate-balanced posterior
and resulting tPoS conditional on a set of weights.
Following the potential outcomes framework, let \(Y^{(z)}\) denote the potential outcome under treatment \(z\in\{0,1\}\). Under the stable unit treatment value assumption (SUTVA) \citep{rubin1974estimating}, the observed outcome satisfies \(Y=Y^{(Z)}=Y^{(1)}Z+Y^{(0)}(1-Z)\). Let \(\pi_z=\Pr(Z=z\mid \boldsymbol{X}, S=0)\in(0,1)\) denote the known phase II randomization probability for treatment \(z\). We make the following assumptions.

\begin{enumerate}

    \item[] Assumption 1: (Randomization)
    $\{Y^{(0)},Y^{(1)}\}\perp\!\!\!\perp Z
    \mid\boldsymbol X,S=0$, and
    $\Pr(Z=z\mid\boldsymbol X,S=0)=\pi_z\in(0,1)$,
    for $z=0,1$. 
    \item[] Assumption 2: (Ignorability and positivity of population membership) $\{Y^{(0)},Y^{(1)}\} \perp\!\!\!\perp S \mid \boldsymbol{X}$, and for all $\boldsymbol{x}$ in the support of $\boldsymbol{X}$ under $S=1$, $0 < \Pr(S=0 \mid \boldsymbol{X}=\boldsymbol{x}) < 1$.
\end{enumerate}
Assumption 1 is guaranteed by the design of the randomized trial, while
Assumption~2 permits the conditional outcome information learned in phase~II
to be transported to the phase~III population.

For \(z\in\{0,1\}\) and any measurable function \(g\) satisfying \(E\{|g(Y^{(z)})|\mid S=1\}<\infty\), these conditions yield the g-formula
\begin{equation}
\begin{aligned}
E\{g(Y^{(z)})\mid S=1\}
&=
\int
E\{g(Y^{(z)})\mid\boldsymbol X=\boldsymbol x,S=1\}
p_1(\boldsymbol x)\,d\boldsymbol x
\\
&=
\int
E\{g(Y^{(z)})\mid\boldsymbol X=\boldsymbol x,S=0\}
p_1(\boldsymbol x)\,d\boldsymbol x
\\
&=
\int
E\{g(Y)\mid Z=z,\boldsymbol X=\boldsymbol x,S=0\}
p_1(\boldsymbol x)\,d\boldsymbol x.
\end{aligned}
\label{eq:gformula}
\end{equation}
The first equality follows from iterated expectation, the second from Assumption 2, and the final equality from Assumption 1 and the relation \(Y=Y^{(Z)}\) under SUTVA. The positivity conditions ensure that the required source-population conditional expectations are defined over the target covariate support.


Equation~\eqref{eq:gformula} identifies target outcome
expectations by averaging phase~II conditional outcome
information over the target covariate distribution
$p_1(\boldsymbol x)$. Let
$r(\boldsymbol x)=p_1(\boldsymbol x)/p_0(\boldsymbol x)$ denote the target-to-source covariate density ratio.
Combining \eqref{eq:gformula} with phase~II randomization,
including the independence of treatment assignment
and baseline covariates, gives
\begin{equation}
E\{g(Y^{(z)})\mid S=1\}
=
\frac{
E\{r(\boldsymbol X)g(Y)\mid Z=z,S=0\}
}{
E\{r(\boldsymbol X)\mid Z=z,S=0\}
}.
\label{eq:transport_ratio}
\end{equation}
Because this identity holds for any integrable function
$g$, taking $g(y)=\mathbb I(y\in B)$ for any measurable
set $B$ identifies the full target arm-specific outcome
distribution $p_1(y\mid z)$, which determines the oracle
PoS in Equation~\eqref{eq:oraclepos}. In particular,
taking $g(y)=y$ identifies the target response probability
for binary endpoints or the target mean for continuous
endpoints, while $g(y)=y^2$ additionally identifies
the target second moment for continuous endpoints.

This identification result motivates using covariate-adjustment weights in the Bayesian analysis of phase~II data. Let $w_i$ denote the raw weight for phase~II participant $i$, constructed using individual- or aggregate-level target information as described in later sections. 
To determine the overall contribution of the weighted phase~II data, we calculate the Kish effective sample size $\tilde n_0=(\sum_{i=1}^{n_0}w_i)^2/ \sum_{i=1}^{n_0}w_i^2$ \citep{kish1965survey} and rescale the weights as $\tilde w_i=w_i\tilde n_0/ \sum_{j=1}^{n_0}w_j$.  
We incorporate this transported evidence through the covariate-balanced posterior
\begin{equation}
p^*(\boldsymbol\varphi\mid D_0)
\propto
\pi_0(\boldsymbol\varphi)
\prod_{i=1}^{n_0}
p(y_{0i}\mid z_{0i},\boldsymbol\varphi)^{\widetilde w_i}.
\label{eq:weighted_posterior}
\end{equation}
Here, $\boldsymbol\varphi$ denotes the parameters
of the marginal outcome model fitted to the
weighted phase~II data for phase~III
prediction. The estimated weights are treated
as fixed when constructing the covariate-balanced
posterior. The population density-ratio identity
in Equation~\eqref{eq:transport_ratio} provides
the theoretical motivation for covariate-adjustment
weighting.

For the Bernoulli marginal model, we assume independent
priors $\pi_0(\theta_z)=\operatorname{Beta}(a,b)$,
$z=0,1$, with small positive $a,b$, such as $a=b=0.5$.
The arm-specific covariate-balanced posterior is
\[
p^*(\theta_z\mid D_0)
=
\operatorname{Beta}\!\left(
a+\sum_{i:z_{0i}=z}\widetilde w_i y_{0i},
\;
b+\sum_{i:z_{0i}=z}\widetilde w_i(1-y_{0i})
\right).
\]

For the normal marginal model, we assume independent
priors $\pi_0(\mu_z,\sigma_z^2)\propto(\sigma_z^2)^{-1}$,
$z=0,1$. The arm-specific covariate-balanced posterior is
\[
p^*(\mu_z,\sigma_z^2\mid D_0)
\propto
(\sigma_z^2)^{-1-\frac{1}{2}\sum_{i:z_{0i}=z}\widetilde w_i}
\exp\!\left\{
-\frac{\sum_{i:z_{0i}=z}\widetilde w_i(y_{0i}-\mu_z)^2}
{2\sigma_z^2}
\right\}.
\]
In both cases, the joint posterior factorizes across treatment arms.

Let $\boldsymbol\varphi=(\theta_0,\theta_1)$ for
binary endpoints and
$\boldsymbol\varphi=(\mu_0,\sigma_0^2,\mu_1,\sigma_1^2)$
for continuous endpoints. We use the fitted
covariate-balanced marginal model for phase~III prediction, where $\theta_z$ denotes
the arm-specific response probability for binary
endpoints, and $\mu_z$ and $\sigma_z^2$ denote
the corresponding mean and variance for
continuous endpoints. Averaging the future-outcome
model over the covariate-balanced posterior gives
the transported posterior predictive distribution:
\begin{align}
p^*(\boldsymbol y_1\mid\boldsymbol z_1,D_0 )
=
\int
p(\boldsymbol y_1\mid\boldsymbol z_1,\boldsymbol\varphi)
p^*(\boldsymbol\varphi\mid D_0 )
\,d\boldsymbol\varphi .
\label{eq:balancedppd}
\end{align}

To estimate the oracle target PoS in
Equation~\eqref{eq:oraclepos}, we replace the
unknown true target outcome distribution
\(p_1(\boldsymbol y_1\mid\boldsymbol z_1)\)
with the covariate-balanced posterior predictive
distribution in Equation~\eqref{eq:balancedppd}.
The resulting transported PoS is
\begin{align}
\operatorname{tPoS}(D_0)
&=
\int\!\!\int
\mathbb I\left(
\mathrm{Success}\mid D_1^{(y,z)}
\right)
p^*(\boldsymbol y_1\mid\boldsymbol z_1,D_0)
p(\boldsymbol z_1)
\,d\boldsymbol y_1\,d\boldsymbol z_1.
\label{eq:balanced}
\end{align}

In the next two subsections, we consider two settings based on the available target
covariate information: (1) individual-level data are available
for both populations; and (2) individual-level data are
available only for phase~II, while phase~III information is
limited to aggregate covariate summaries.

\subsubsection{tPoS with Individual-Level Target Data} \label{covariatebalance:individual}


In some applications, it is feasible to sample patients who meet the eligibility criteria for the phase III target population from electronic health records, disease registries, or historical standard-of-care cohorts to construct a target dataset containing individual-level covariate information. In such settings, we use inverse-odds weighting (IOW, \cite{westreich2017transportability}) to adjust for covariate-distribution differences between the phase~II source and target phase~III populations  
 for estimating the PoS.

Specifically, let $e(\boldsymbol{X})=\Pr(S=1\mid\boldsymbol{X})$ denote the conditional probability of belonging to the target phase III population. The population IOW is defined as
\begin{align}
    w^{IO}(\boldsymbol{X})=\frac{e(\boldsymbol{X})}{1-e(\boldsymbol{X})}=\frac{\Pr(S=1)p_1(\boldsymbol{X})}{\Pr(S=0)p_0(\boldsymbol X)} = \frac{\Pr(S=1)}{\Pr(S=0)}r(\boldsymbol{X}),
\end{align}
so $w^{IO}(\boldsymbol{X})$ is proportional to the density ratio $r(\boldsymbol{X})$ introduced in Section \ref{general}. By Equation~\eqref{eq:transport_ratio},
population IOW therefore identifies
the target phase~III arm-specific potential-outcome
distributions under Assumptions~1--2. The identification and consistency properties of IOW for transporting trial-based causal quantities are well established under standard transportability and sampling-score estimation assumptions \citep{dahabreh2020extending}.

Flexible machine-learning methods have been investigated
for propensity-score estimation \citep{lee2010improving,chang2022flexible}.
In our implementation, we use Bayesian additive regression
trees (BART). 
For phase~II participants, writing
$\boldsymbol x_i=\boldsymbol x_{0i}$ for simplicity, the estimated
IOW is
\(
w_i^{IO}=\hat e(\boldsymbol x_i)/
[1-\hat e(\boldsymbol x_i)].
\) 

To mitigate instability from extreme weights, symmetric
propensity score trimming \citep{crump2009dealing}
can be applied using fixed cutoffs, for example, by 
excluding observations with
$\hat e(\boldsymbol x_i)\notin[0.05,0.95]$.
The resulting $w_i^{IO}$ is used as the raw weight $w_i$
in the covariate-balanced posterior construction in
Section~\ref{general}.

\subsubsection{{tPoS with Aggregate-Level Target Data}}\label{covariatebalance:aggregate}

A more common setting is that individual-level data from the target phase III population are unavailable because the trial has not yet been conducted. Instead, the target population is characterized only by aggregate-level data (AGD), such as covariate means, standard deviations, and proportions, obtained from published studies, previous trials involving similar populations, disease registries, or electronic health records. To accommodate this setting, we adopt the method-of-moments (MoM) weighting approach \citep{yan2025target},
which assigns weights to phase II participants
so that their weighted covariate summaries
match those reported for the target phase III
population.

To incorporate these aggregate-level summaries, we select a set of functions $h_j(\boldsymbol x)$ of the baseline covariates $\boldsymbol x$, chosen to capture relevant characteristics of the covariate distribution, such as means, variances, and quantiles.
Let
$\boldsymbol h(\boldsymbol x)
=(h_1(\boldsymbol x),\ldots,h_q(\boldsymbol x))^\top$
denote the vector of $q$ selected functions. 
We define $\boldsymbol h_1
=E\{\boldsymbol h(\boldsymbol X)\mid S=1\}$
as the corresponding expectations in the target phase III population, and let $\bar{\boldsymbol h}_1$
denote their estimates obtained directly from or derived using the available AGD.

For illustration, consider a single baseline covariate \(x\) representing age. If the phase III AGD report a mean age of 60 years, we choose \(\boldsymbol h(x)=x\) and set \(\bar{\boldsymbol h}_1=60\). If both the mean and standard deviation of age are reported as 60 and 10 years, respectively, we choose \(\boldsymbol h(x)=(x,x^2)^\top\) and set \(\bar{\boldsymbol h}_1=(60,60^2+10^2)^\top=(60,3700)^\top\), corresponding to the first two raw moments.

Alternatively, suppose the phase III AGD report age-group proportions of 0.22, 0.48, and 0.30 for ages 40–49, 50–59, and 60–69 years, respectively. We can then choose
\[
\boldsymbol h(x)=
\begin{pmatrix}
\mathbb I\{40\leq x<50\}\\
\mathbb I\{50\leq x<60\}
\end{pmatrix},
\qquad
\bar{\boldsymbol h}_1=
\begin{pmatrix}
0.22\\
0.48
\end{pmatrix},
\]
where \(\mathbb I\{\cdot\}\) denotes the indicator function. Because these age groups are mutually exclusive and exhaustive, matching the first two proportions automatically ensures that the third is matched.
Similarly, for a binary covariate \(x\) representing sex, we can choose \(\boldsymbol h(x)=\mathbb I\{x=\mathrm{female}\}\) and set \(\bar{\boldsymbol h}_1\) equal to the proportion of female patients reported in the AGD.

Following the exponential-tilting framework commonly considered in AGD-based transportability methods \citep{josey2021transporting,sheng2026transportable}, we assume that the source-to-target covariate density ratio satisfies
\begin{align}
\log r(\boldsymbol X)
=c+\boldsymbol h(\boldsymbol X)^\top
\boldsymbol\eta,
\label{eq:agd_ratio}
\end{align}
where $c$ is a normalizing constant and
$\boldsymbol\eta\in\mathbb R^q$ is a coefficient vector. This assumption
is equivalent to a logistic population-membership
model with log-odds linear in
$\boldsymbol h(\boldsymbol X)$. Under this specification, the population
MoM weight is defined as
\[
w^M(\boldsymbol X)
=\exp\{\boldsymbol h(\boldsymbol X)^\top
\boldsymbol\eta\}
\propto r(\boldsymbol X).
\]

The following proposition establishes the
population-level moment-matching and causal
identification properties of these weights.

\begin{proposition}[Identification of target outcome distributions and oracle PoS]
\label{prop:agd_mom}
\itshape

Suppose Assumptions~1--2 and
Equation~\eqref{eq:agd_ratio} hold.
Then the MoM-weighted phase II covariate moments match the corresponding target moments:
\[
\frac{
E[w^M(\boldsymbol X)\boldsymbol h(\boldsymbol X)
\mid S=0]
}{
E[w^M(\boldsymbol X)\mid S=0]
}
=\boldsymbol h_1.
\]
If $\operatorname{Var}\{\boldsymbol h(\boldsymbol X)
\mid S=1\}$ is positive definite,
$\boldsymbol\eta$ is the unique coefficient
vector for which exponential-tilting weights
reproduce $\boldsymbol h_1$, among coefficients
for which the required expectations are
well-defined. 
Furthermore, for each $z\in\{0,1\}$ and
any measurable function $g$ satisfying
$E\{|g(Y^{(z)})|\mid S=1\}<\infty$,
\begin{align}
E\{g(Y^{(z)})\mid S=1\}
&=
\frac{
E[w^M(\boldsymbol X)g(Y)\mid Z=z,S=0]
}{
E[w^M(\boldsymbol X)\mid Z=z,S=0]
}.
\label{eq:agd_identification}
\end{align}
Thus, the resulting weights identify the target arm-specific potential-outcome distributions from the phase II observed-data distribution and the target covariate moments $\boldsymbol h_1$. Consequently, under the independent-participant
sampling model used in
Equation~\eqref{eq:oraclepos}, with the
phase III sample size, randomization scheme,
and success criterion prespecified,
$\operatorname{PoS}_1^{\mathrm{oracle}}$
is also identifiable.

\end{proposition}

The proof is provided in Supplementary
Section~{S2.1}. In practice, $\boldsymbol\eta$ is unknown and
must be estimated using the phase II data and
the available phase III AGD. For phase II participant
$i=1,\ldots,n_0$, write
$\boldsymbol x_i=\boldsymbol x_{0i}$
for simplicity. 
We obtain $\hat{\boldsymbol\eta}$ by matching
the weighted phase II covariate summaries
to the corresponding target phase III summaries:
\begin{align}
\frac{
\sum_{i=1}^{n_0}
\boldsymbol h(\boldsymbol x_i)
\exp\{\boldsymbol h(\boldsymbol x_i)^\top
\hat{\boldsymbol\eta}\}
}{
\sum_{i=1}^{n_0}
\exp\{\boldsymbol h(\boldsymbol x_i)^\top
\hat{\boldsymbol\eta}\}
}
=\bar{\boldsymbol h}_1.
\label{momequation}
\end{align}
The resulting raw MoM weights are
\(
w_i^M
=\exp\{\boldsymbol h(\boldsymbol x_i)^\top
\hat{\boldsymbol\eta}\}.
\)

To solve Equation~\eqref{momequation}, let
$\boldsymbol d_i=\boldsymbol h(\boldsymbol x_i)
-\bar{\boldsymbol h}_1$.
For a candidate coefficient vector
$\boldsymbol\eta$, the moment-matching equation
can be rewritten as
$\sum_{i=1}^{n_0}\boldsymbol d_i
\exp\{\boldsymbol d_i^\top\boldsymbol\eta\}
=\boldsymbol 0$.
This is the gradient of
$Q(\boldsymbol\eta)=\sum_{i=1}^{n_0}
\exp\{\boldsymbol d_i^\top\boldsymbol\eta\}$.
The Hessian matrix is
$\nabla^2Q(\boldsymbol\eta)
=\sum_{i=1}^{n_0}
\exp\{\boldsymbol d_i^\top\boldsymbol\eta\}
\boldsymbol d_i\boldsymbol d_i^\top
\succeq 0$.
Thus, $Q(\boldsymbol\eta)$ is convex,
and solving Equation~\eqref{momequation}
is equivalent to minimizing
$Q(\boldsymbol\eta)$.

The consistency of $\hat{\boldsymbol\eta}$ and
the resulting MoM-weighted estimators of target
arm-specific outcome expectations is established
under the regularity conditions given in
Supplementary Section~{S2.2}. 
As with propensity score weighting, trimming
may be applied to mitigate instability from
extreme MoM weights \citep{yan2025target}.
The resulting weights are used as the raw
weights $w_i$ in the covariate-balanced
posterior construction in Section~\ref{general}.

\section{Simulation Study} \label{Simulation}

\subsection{Set-up} \label{setup}

We evaluated the proposed tPoS methods through simulation studies with binary and continuous endpoints. For binary endpoints, we considered a phase~II trial with \(n_0=100\) participants (50 per arm) and independent noninformative priors
\(
\pi(\theta_z)=\operatorname{Beta}(0.5,0.5),~z=0,1.
\)
The future phase~III trial uses the Bayesian decision rule in \eqref{successdef} with \(\delta_0=0\) and the same prior specification. A total of 400 patients (200 per arm) are enrolled to achieve 90\% power for a prespecified treatment effect \(\Delta=\theta_1-\theta_0=0.16\), with \(\theta_1=0.76\) and \(\theta_0=0.60\). The decision threshold is calibrated to \(\gamma=0.9774\) to control the one-sided Type~I error at 0.025.

For continuous endpoints, we considered a phase~II trial with \(n_0=100\) participants (50 per arm) and independent Jeffreys priors
\(
\pi(\mu_z,\sigma_z^2)\propto(\sigma_z^2)^{-1},~ z=0,1.
\)
The future phase~III trial uses the Bayesian decision rule in \eqref{successdef} with \(\delta_0=0\) and the same prior specification. A total of 400 patients (200 per arm) are enrolled to achieve 90\% power for a prespecified treatment effect \(\Delta=\mu_1-\mu_0=0.33\), with \(\mu_1=0.93\), \(\mu_0=0.60\), and \(\sigma_0^2=\sigma_1^2=1\). The decision threshold is calibrated to \(\gamma=0.978\) to control the one-sided Type~I error at 0.025.

In each simulation replicate, we generate a phase~II dataset consisting of individual-level covariates and outcomes, as well as phase III target population datasets, which can be individual-level covariate data or only aggregate covariate summaries. 

Specifically, for the phase II data, we consider $p=2$ independent covariates: one binary covariate, $X_1$, and one continuous covariate, $X_2$, generated from 
\begin{align}
X_1 &\sim \mathrm{Bernoulli}(p_{X_1}), \qquad
X_2 \sim \lambda \mathcal{N}(\mu_{X_2,1}, \sigma_{X_2,1}^2)
      +(1-\lambda)\mathcal{N}(\mu_{X_2,2}, \sigma_{X_2,2}^2).\label{p2distribution}
\end{align}
For the phase~III population, the covariates $X_1, X_2$ are from
\begin{align}
X_1 &\sim \mathrm{Bernoulli}(p_{X_1}^*), \qquad
X_2 \sim \lambda^* \mathcal{N}(\mu_{X_2,1}^*, \sigma_{X_2,1}^{*2})
      +(1-\lambda^*)\mathcal{N}(\mu_{X_2,2}^*, \sigma_{X_2,2}^{*2}).
\label{p3distribution}
\end{align}

For the binary endpoint, given the covariates $(X_1,X_2)$ and the treatment assignment $Z_i$, $y_i$ is generated from the logit model: 
\begin{equation}
\begin{aligned}
\operatorname{logit}\{\Pr(Y_i=1 \mid X_{1,i},X_{2,i},Z_i)\}
=&\ \beta_0
+\beta_{1}X_{1,i}
+\beta_{2}X_{2,i}
+\beta_{2}^{'}\,\mathbb{I}(X_{2,i}\le 0)+\beta_{\mathrm{trt}} Z_i \\
&
+\beta_{\mathrm{trt},1}\, Z_i X_{1,i}
+\beta_{\mathrm{trt},2}\, Z_i X_{2,i}
+\beta_{\mathrm{trt},2}^{'}\, Z_i\,\mathbb{I}(X_{2,i}\le 0).
\end{aligned}
\label{binaryoutcome}
\end{equation}
Varying the values of $(p_{X_1}^*,\mu_{X_2,1}^*,{\sigma_{X_2,1}^*}^2, \mu_{X_2,2}^*,{\sigma_{X_2,2}^*}^2,\lambda^*)$, $(p_{X_1},\mu_{X_2,1},{\sigma_{X_2,1}}^2,\mu_{X_2,1},{\sigma_{X_2,1}}^2,\lambda)$, and $(\beta_0,\beta_{X_1},\beta_{X_2},\beta_{X_2}',\beta_{\text{trt}},\beta_{\text{trt},1},\beta_{\text{trt},2},\beta_{\text{trt},2}')$, we consider a total of seven simulation scenarios, as shown in Tables {S1} and {S2}. 

For the continuous endpoint, given the covariates $(X_1,X_2)$ and the treatment assignment $Z_i$, $y_i$ is generated from the following normal model:
\begin{equation}
\begin{aligned}
y_i =&\beta_0+ \beta_{1} X_{1,i} + \beta_{2} X_{2,i}+\beta_{2}^{'}\,\mathbb{I}(X_{2,i}\le 0)+ \beta_{\text{trt}} Z_i
       \\ &+ \beta_{\text{trt},1} Z_i X_{1,i}
         + \beta_{\text{trt},2} Z_i X_{2,i}+\beta_{\mathrm{trt},2}^{'}\, Z_i\,\mathbb{I}(X_{2,i}\le 0)
         + \epsilon_i ,
\end{aligned}
\label{contoutcome}
\end{equation}
where $\epsilon_i\sim N(0,1)$. Similarly to the binary endpoints, we also consider a total of {seven scenarios}, given in Tables S3 and S4.

We compared the proposed tPoS method with two alternative approaches. The first, referred to as Standard, represents the conventional PoS calculation that assumes no covariate shift between the phase~II and phase~III populations. It calculates PoS directly from the generated phase~II dataset without adjusting for differences in covariate distributions. The second, NoShift, serves as an idealized benchmark. It uses the same PoS calculation as Standard but is evaluated under a data-generating scenario where no covariate shift exists. 

We considered three variants of tPoS. 
The first variant, $\text{tPoS}_\text{IPD}$, uses individual-level phase~III covariates from a simulated historical sample of size 200 generated under \eqref{p3distribution}. The second variant, $\text{tPoS}_\text{AGD}$, uses only the sample means of $X_1$ and $X_2$ and the sample variance of $X_2$. The third variant, $\text{tPoS}_\text{AGD-Q}$, additionally incorporates the known phase~III probabilities of $X_2$ falling within the intervals $(-\infty,-0.174]$, $(-0.174,0.5]$, $(0.5,1.174]$, and $(1.174,\infty)$, with cutpoints derived from standard normal quantiles. This variant illustrates the use of grouped distributional summaries in addition to aggregate moments.

We summarized the performance of each method using the mean and the 25\%, 50\%, and 75\% quantiles of the calculated PoS across simulation replicates. To further evaluate the performance of PoS in guiding go/no-go decisions, we considered
\(
C_\alpha = \Pr\bigl(\operatorname{PoS} \le \phi_\alpha\bigr),
\) 
where $\phi_\alpha$ denotes the lower $\alpha$-quantile of the PoS distribution under NoShift. Under a decision rule that declares ``no-go'' when $\operatorname{PoS} \le \phi_\alpha$, $C_\alpha$ represents the corresponding no-go probability for each method in a given simulation scenario. By construction, $C_\alpha=\alpha$ for NoShift, providing a benchmark against which the other methods are evaluated. Values of $C_\alpha$ closer to $\alpha$ indicate better agreement with NoShift in terms of the no-go decision rate.

\subsection{Results} \label{result}
Table~\ref{table:resultsbinary} summarizes the results for the binary-endpoint based on 2000 simulations.  
Compared with Standard, tPoS provides more accurate PoS estimates in the presence of covariate shift between phase~II and phase~III populations, yielding estimates closer to the NoShift benchmark, while maintaining comparable performance in the absence of covariate shift.

Specifically, in Scenario~1, where there is no covariate shift, all methods yield similar PoS estimates, with medians around 0.800 and means around 0.730. The tPoS methods also produce $C_\alpha$ values close to $\alpha$ for $\alpha\in{0.20,0.30,0.40}$, indicating decision behavior comparable to Standard and NoShift. The small differences of tPoS from Standard reflect the reweighting for covariate balancing and its associated reduction in effective sample size.

When covariate shift is present (Scenarios~2--7), the improvement from tPoS depends on how well the available phase~III target information captures the relevant features of the shift. When these features are adequately represented in the balancing procedure, tPoS produces PoS estimates closer to NoShift and improves decision calibration relative to Standard.

In Scenarios~2--5, the covariate shifts are characterized by changes in the mean and/or variance and can therefore be captured by the available balancing information. Standard can substantially over- or underestimate PoS, whereas all tPoS methods considerably reduce these discrepancies. For example, in Scenario~3, Standard overestimates the median PoS (0.805 vs.\ 0.562 under NoShift), while the medians for $\mathrm{tPoS}_{\mathrm{IPD}}$, $\mathrm{tPoS}_{\mathrm{AGD}}$, and $\mathrm{tPoS}_{\mathrm{AGD\text{-}Q}}$ are 0.554, 0.536, and 0.540, respectively. Conversely, in Scenario~4, Standard underestimates the median PoS (0.527 vs.\ 0.828 under NoShift), whereas the corresponding the medians for $\mathrm{tPoS}_{\mathrm{IPD}}$, $\mathrm{tPoS}_{\mathrm{AGD}}$, and $\mathrm{tPoS}_{\mathrm{AGD\text{-}Q}}$ are 0.763, 0.781, and 0.778. The tPoS methods also yield $C_\alpha$ values consistently closer to $\alpha$. Among the tPoS methods, $\mathrm{tPoS}_{\mathrm{AGD}}$ performs best in Scenarios~2--5 because the covariate shifts are fully characterized by the first two moments, which it directly aligns. Although $\mathrm{tPoS}_{\mathrm{IPD}}$ is more flexible, subject-level reweighting may sacrifice efficiency, whereas the additional quantile constraints in $\mathrm{tPoS}_{\mathrm{AGD\text{-}Q}}$ may similarly reduce efficiency when they provide little information beyond the first two moments.

Scenarios~6--7 instead involve structural changes in the mixture distribution of $X_2$ that are not reflected in its mean and variance. Consequently, balancing the first two moments alone is insufficient, and methods using richer distributional information show greater improvement. In Scenario~7, for example, Standard overestimates the median PoS (0.801 vs.\ 0.524 under NoShift). $\mathrm{tPoS}_{\mathrm{AGD}}$ reduces the median to 0.646, whereas $\mathrm{tPoS}_{\mathrm{IPD}}$ and $\mathrm{tPoS}_{\mathrm{AGD\text{-}Q}}$ yield 0.533 and 0.515, respectively. The same pattern is observed for decision calibration, with the latter two methods producing $C_\alpha$ values substantially closer to $\alpha$ than Standard. 
Scenario~6 provides a more direct illustration because the covariate shift is purely structural, with no differences in the first two moments between the phase~II and phase~III populations. As a result, $\mathrm{tPoS}_{\mathrm{AGD}}$ cannot adequately detect the shift: its median PoS is 0.553, only slightly different from 0.544 for Standard, compared with 0.748 under NoShift. In contrast, $\mathrm{tPoS}_{\mathrm{AGD\text{-}Q}}$ and $\mathrm{tPoS}_{\mathrm{IPD}}$ yield medians of 0.671 and 0.696, respectively, demonstrating the benefit of incorporating information beyond the first two moments when the covariate distribution changes structurally.

Table~\ref{table:resultscontinuous} presents the simulation results for continuous endpoints.
The overall patterns are consistent with those observed for the binary endpoint. 

\subsection{Sensitivity Analysis} \label{sensitivity}

The proposed tPoS framework assumes that the phase~III baseline covariate distribution is available, either as individual patient data or summary statistics. To evaluate the robustness of tPoS to inaccuracies in this information, we conducted sensitivity analyses examining the impact of misspecification of the phase~III covariate distribution used in tPoS estimation.
The simulated individual-level phase~III covariates continued to follow \eqref{p3distribution}, but the target dataset used to construct the balancing weights and calculate tPoS was generated with
$X_1\sim\mathrm{Bernoulli}(p_{X_1}^*-0.05)$,
while the distribution of $X_2$ remained unchanged. Thus, the covariate information supplied to the tPoS procedure differed moderately from the true phase~III population distribution, with all other simulation settings held fixed. 

\begin{figure}[!htbp]
  \centering
  \includegraphics[width=\textwidth]{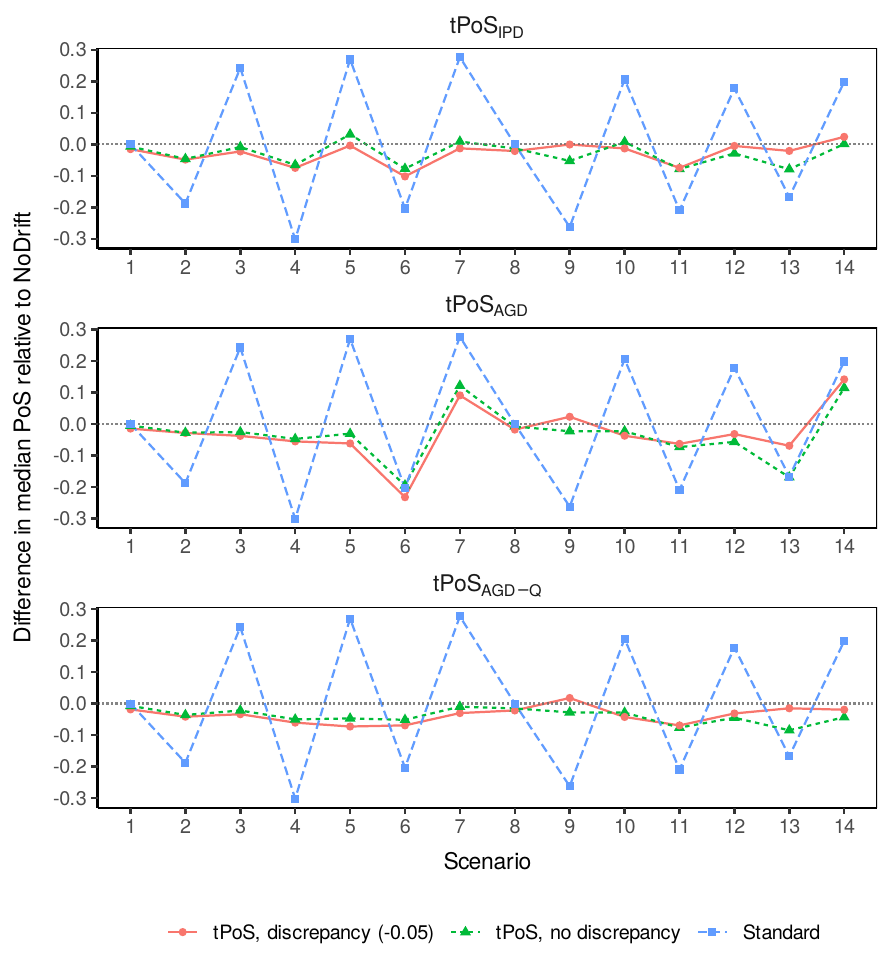}
  \caption{Sensitivity analysis when phase III covariate distribution is misspecified. 
  The grey dotted line indicates the value of 0.}
  \label{fig:Fig-2}
\end{figure}

As shown in Figure~\ref{fig:Fig-2}, the resulting tPoS differs from those obtained under correctly specified phase III covariate distributions.
The magnitude and direction of these deviations vary across scenarios, depending on the role of $X_1$ in the outcome model and the underlying phase~II--III covariate shift. Nevertheless, the discrepancies are generally modest, and the resulting tPoS remains substantially closer to NoShift than that from the unadjusted Standard approach. The impact is relatively small for $\mathrm{tPoS}{_\mathrm{IPD}}$ and $\mathrm{tPoS}_{\mathrm{AGD\text{-}Q}}$, whereas somewhat greater variation is observed for $\mathrm{tPoS}_{\mathrm{AGD}}$ in several scenarios.

\section{Application} \label{application}
We applied the tPoS approach to the published MUSE--TULIP-1 trial pair \citep{furie2017anifrolumab,furie2019type}. MUSE was a randomized, double-blind, placebo-controlled Phase~IIb trial evaluating intravenous anifrolumab 300~mg or 1,000~mg administered every four weeks, in addition to standard therapy, in patients with moderate-to-severe systemic lupus erythematosus \citep{furie2017anifrolumab}. The modified intention-to-treat population included 305 patients. The primary endpoint was SRI-4 response at Week~24 with a sustained reduction in oral corticosteroid use. In the 300-mg group, this endpoint was achieved by 34.3\% of patients, compared with 17.6\% in the placebo group ($p=0.014$). At Week~52, SRI-4 response 
was achieved by 62 of 99 patients (62.6\%) receiving anifrolumab 300~mg and 41 of 102 patients (40.2\%) receiving placebo (nominal $p<0.001$). 

Motivated by the positive results of MUSE, a randomized, double-blind, placebo-controlled Phase~III trial, TULIP-1, was conducted to evaluate anifrolumab 300 mg 
versus placebo, in addition to standard therapy, in the same disease setting \citep{furie2019type}. The primary endpoint was the between-group difference in the proportion of patients achieving an SRI-4 response at Week~52 with anifrolumab versus placebo. For this primary comparison, the planned sample size was 360 patients, with 180 patients per group. The trial did not meet its primary endpoint: response was observed in 36.1\% of the patients receiving anifrolumab and 40.2\% of the patients receiving placebo ($p=0.41$).

For the tPoS application, we used the anifrolumab 300-mg and placebo groups from MUSE as the Phase~II source population and the corresponding groups from TULIP-1 to define the Phase~III target population. The key baseline characteristics of the MUSE source population and the TULIP-1 target population are summarized in Table~\ref{tab:muse_tulip1}. The most pronounced difference between the two populations was in racial composition, with white patients accounting for 37.8\% of the MUSE population and 72.0\% of the TULIP-1 population. Smaller difference was observed in the proportion of patients with a high type I IFN gene signature (75.1\% versus 82.1\%), whereas the mean age (39.2 versus 41.5 years), the mean SLEDAI-2K score (10.9 versus 11.4), and the proportion receiving an oral corticosteroid dose of at least 10~mg/day at baseline (59.2\% versus 56.3\%) were relatively more similar.

We applied both the standard PoS approach and the proposed $\mathrm{tPoS}_{\mathrm{AGD}}$ 
to the trial data. Because individual participant data from MUSE were unavailable, we reconstructed semi-synthetic Phase~II data using the published arm-specific marginal summaries. For the target phase III population, we used the summary statistics of covariates reported in TULIP-1. The results are reported in Table~\ref{tab:muse_tulip1}. While the standard PoS yielded an overly optimistic estimate of 0.923, transporting the Phase~II evidence to the target TULIP-1 population via $\text{tPoS}_{\text{AGD}}$ adjusted the projected PoS downward to 0.433. This substantial divergence underscores the critical importance of accounting for covariate shift between trial populations. By providing a more realistic assessment, the proposed tPoS methodology can prompt investigators to exercise greater caution when evaluating whether a phase~III trial is warranted, encourage more deliberate definition of the confirmatory target population, and ultimately reduce the risk of late-stage trial failure.

\section{Discussion} \label{discuss}

By adopting a transportability perspective, we have proposed a Bayesian covariate-balanced framework for PoS evaluation when baseline characteristics differ between phase~II and phase~III populations. 
This framework formally bridges evidence generated in a source cohort with an anticipated target population, adjusting the phase~II posterior distribution to account for covariate shift prior to launching the confirmatory study. By leveraging available phase~III covariate data, at either the individual or aggregate level, the method constructs balancing constraints that calibrate phase~II findings to the intended phase~III population. Our numerical studies demonstrate that the proposed tPoS approach yields a substantially more accurate and dependable assessment of trial success than the standard PoS method.

We introduced the framework for a randomized phase~II trial. However, some phase~II trials are single-arm.  
In such cases, real world data or historical data can be used to construct an external control. For example, propensity-score methods can identify comparable external patients for incorporation into the single-arm study \citep{jiang2023basic}.
The resulting phase~II evidence can then be aligned with the anticipated phase~III target population using the covariate-balancing framework in Section~\ref{adjust}. 

In some settings, phase~III target population information may not be available at the planning stage. In such cases, a deferred PoS strategy can be adopted, whereby PoS is updated as partial phase~III baseline characteristics become available. If outcome data from the partially enrolled phase~III cohort are also observed (e.g., at interim analysis), a Bayesian adaptive borrowing approach, such as the self-adapting mixture (SAM) prior \citep{yang2023SAM} may be further employed to dynamically integrate covariate-balanced phase~II evidence with accumulating phase~III data for PoS evaluation. 
In addition, our framework assumes that the phase II study provides treatment-effect information on the endpoint used to evaluate phase III success. In practice, however, phase II development may rely primarily on a short-term, intermediate, or surrogate endpoint, whereas phase III success is evaluated using a different clinical endpoint. In such settings, an additional model is needed to translate the phase II treatment effect into a predictive distribution for the phase III endpoint before PoS can be evaluated \citep{quan2023utilization}. 
Future work could integrate such surrogate- or endpoint-bridging approaches with the proposed covariate-balanced framework to accommodate both population and endpoint differences between phase II and phase III.

\bibliography{citation}
\clearpage

\begin{table}[!htbp]
\caption{Simulation results for a binary endpoint,  including
25\%, 50\%, and 75\% quantiles (denoted as Q25, Q50, and Q75, respectively) and mean of PoS, and  
$C_\alpha = \Pr(\operatorname{PoS} \le \phi_\alpha)$, where $\phi_\alpha$ is the lower $\alpha$-quantile for NoShift.}
\label{table:resultsbinary}
    \centering
    \small
    \begin{tabular}{cccccccccc}
    \toprule
        \textbf{Scenario} & \textbf{Description} & \textbf{Method} & \textbf{Q25} & \textbf{Q50} & \textbf{Q75} & \textbf{Mean} & $\mathbf{C}_\mathbf{0.20}$ & $\mathbf{C}_\mathbf{0.30}$ & $\mathbf{C}_\mathbf{0.40}$\\
    \midrule
    
\multirow{5}{*}{1}  &  \multirow{5}{*}{No covariate shift}          & NoShift & 0.609 & 0.805 & 0.927 & 0.739 & 0.200 & 0.300 & 0.400 \\
  &  & Standard & 0.609 & 0.805 & 0.927 & 0.739 & 0.200 & 0.300 & 0.400 \\
  &  & $\text{tPoS}_\text{IPD}$ & 0.587 & 0.798 & 0.925 & 0.732 & 0.214 & 0.321 & 0.400 \\
  &  & $\text{tPoS}_\text{AGD}$ & 0.597 & 0.800 & 0.925 & 0.734 & 0.216 & 0.303 & 0.394 \\
  &  & $\text{tPoS}_\text{AGD-Q}$ & 0.584 & 0.798 & 0.924 & 0.730 & 0.224 & 0.321 & 0.408 \\
  \\
\multirow{5}{*}{2}  & \multirow{5}{*}{\shortstack{Mean shift from\\Phase II to Phase III}}           & NoShift & 0.545 & 0.750 & 0.904 & 0.696 & 0.200 & 0.300 & 0.400 \\
  &            & Standard & 0.323 & 0.562 & 0.775 & 0.546 & 0.405 & 0.544 & 0.634 \\
  &            & $\text{tPoS}_\text{IPD}$ & 0.443 & 0.704 & 0.883 & 0.645 & 0.276 & 0.380 & 0.472 \\
  &            & $\text{tPoS}_\text{AGD}$ & 0.467 & 0.722 & 0.895 & 0.660 & 0.263 & 0.363 & 0.449 \\
  &            & $\text{tPoS}_\text{AGD-Q}$ & 0.456 & 0.714 & 0.892 & 0.656 & 0.270 & 0.372 & 0.457 \\
  \\
\multirow{5}{*}{3}  &  \multirow{5}{*}{\shortstack{Mean shift from\\Phase II to Phase III}}  & NoShift & 0.323 & 0.562 & 0.775 & 0.546 & 0.200 & 0.300 & 0.400 \\
  &            & Standard & 0.609 & 0.805 & 0.927 & 0.739 & 0.055 & 0.099 & 0.148 \\
  &            & $\text{tPoS}_\text{IPD}$ & 0.341 & 0.554 & 0.768 & 0.546 & 0.186 & 0.287 & 0.405 \\
  &            & $\text{tPoS}_\text{AGD}$ & 0.320 & 0.536 & 0.746 & 0.531 & 0.200 & 0.310 & 0.420 \\
  &            & $\text{tPoS}_\text{AGD-Q}$ & 0.322 & 0.540 & 0.754 & 0.531 & 0.197 & 0.300 & 0.423 \\
  \\
 \multirow{5}{*}{4} &  \multirow{5}{*}{\shortstack{Mean and variance shift\\ from Phase II to Phase III}}   & NoShift & 0.637 & 0.828 & 0.942 & 0.761 & 0.201 & 0.300 & 0.400 \\
  &            & Standard & 0.302 & 0.527 & 0.747 & 0.518 & 0.562 & 0.678 & 0.783 \\
  &            & $\text{tPoS}_\text{IPD}$ & 0.516 & 0.763 & 0.918 & 0.696 & 0.303 & 0.398 & 0.508 \\
  &            & $\text{tPoS}_\text{AGD}$ & 0.536 & 0.781 & 0.925 & 0.707 & 0.289 & 0.381 & 0.488 \\
  &            & $\text{tPoS}_\text{AGD-Q}$ & 0.541 & 0.778 & 0.918 & 0.703 & 0.295 & 0.384 & 0.490 \\
  \\
\multirow{5}{*}{5}  &  \multirow{5}{*}{\shortstack{Mean and variance shift\\ from Phase II to Phase III}} & NoShift & 0.320 & 0.559 & 0.766 & 0.541 & 0.201 & 0.301 & 0.400 \\
  &            & Standard & 0.637 & 0.828 & 0.942 & 0.761 & 0.049 & 0.076 & 0.125 \\
  &            & $\text{tPoS}_\text{IPD}$ & 0.341 & 0.591 & 0.807 & 0.567 & 0.184 & 0.280 & 0.373 \\
  &            & $\text{tPoS}_\text{AGD}$ & 0.286 & 0.528 & 0.774 & 0.522 & 0.234 & 0.338 & 0.440 \\
  &            & $\text{tPoS}_\text{AGD-Q}$ & 0.277 & 0.512 & 0.765 & 0.518 & 0.245 & 0.348 & 0.454 \\
  \\
\multirow{5}{*}{6}  & \multirow{5}{*}{\shortstack{Distributional shape shift\\from Phase II to Phase III}} & NoShift & 0.527 & 0.748 & 0.899 & 0.691 & 0.200 & 0.300 & 0.400 \\
  &            & Standard & 0.322 & 0.544 & 0.769 & 0.538 & 0.418 & 0.539 & 0.637 \\
  &            & $\text{tPoS}_\text{IPD}$ & 0.422 & 0.671 & 0.858 & 0.623 & 0.296 & 0.400 & 0.496 \\
  &            & $\text{tPoS}_\text{AGD}$ & 0.323 & 0.553 & 0.774 & 0.539 & 0.415 & 0.537 & 0.634 \\
  &            & $\text{tPoS}_\text{AGD-Q}$ & 0.450 & 0.696 & 0.871 & 0.641 & 0.271 & 0.370 & 0.469 \\
  \\
\multirow{5}{*}{7}  & \multirow{5}{*}{\shortstack{Distributional shape shift\\from Phase II to Phase III}} & NoShift & 0.290 & 0.524 & 0.741 & 0.516 & 0.201 & 0.302 & 0.400 \\
  &            & Standard & 0.602 & 0.801 & 0.935 & 0.739 & 0.039 & 0.082 & 0.120 \\
  &            & $\text{tPoS}_\text{IPD}$ & 0.289 & 0.533 & 0.771 & 0.525 & 0.196 & 0.306 & 0.392 \\
  &            & $\text{tPoS}_\text{AGD}$ & 0.407 & 0.646 & 0.846 & 0.609 & 0.120 & 0.198 & 0.284 \\
  &            & $\text{tPoS}_\text{AGD-Q}$ & 0.275 & 0.515 & 0.753 & 0.512 & 0.210 & 0.324 & 0.419\\

    \bottomrule
    \end{tabular}
\end{table}

\begin{table}[!htbp]
\caption{Simulation results for a continuous endpoint, including
25\%, 50\%, and 75\% quantiles (denoted as Q25, Q50, and Q75, respectively) and mean of PoS, and  
$C_\alpha = \Pr(\operatorname{PoS} \le \phi_\alpha)$,
where $\phi_\alpha$ is the lower $\alpha$-quantile for NoShift.}
\label{table:resultscontinuous}
    \centering
    \small
    \begin{tabular}{cccccccccc}
    \toprule
        \textbf{Scenario} & \textbf{Description} & \textbf{Method} & \textbf{Q25} & \textbf{Q50} & \textbf{Q75} & \textbf{Mean} & $\mathbf{C}_\mathbf{0.20}$ & $\mathbf{C}_\mathbf{0.30}$ & $\mathbf{C}_\mathbf{0.40}$\\
    \midrule
        \multirow{5}{*}8  & \multirow{5}{*}{No covariate shift}          & NoShift & 0.634 & 0.835 & 0.940 & 0.760 & 0.200 & 0.300 & 0.400 \\
  &            & Standard & 0.634 & 0.835 & 0.940 & 0.760 & 0.200 & 0.300 & 0.400 \\
  &           & $\text{tPoS}_\text{IPD}$ & 0.621 & 0.822 & 0.933 & 0.752 & 0.213 & 0.315 & 0.421 \\
  &            & $\text{tPoS}_\text{AGD}$ & 0.614 & 0.826 & 0.938 & 0.753 & 0.215 & 0.317 & 0.411 \\
  &            & $\text{tPoS}_\text{AGD-Q}$ & 0.611 & 0.819 & 0.932 & 0.747 & 0.219 & 0.326 & 0.425 \\
  \\
\multirow{5}{*}{9}  & \multirow{5}{*}{\shortstack{Mean shift from\\Phase II to Phase III}} & NoShift & 0.572 & 0.775 & 0.908 & 0.714 & 0.200 & 0.300 & 0.401 \\
&            & Standard & 0.286 & 0.512 & 0.727 & 0.507 & 0.501 & 0.635 & 0.727 \\
  &            & $\text{tPoS}_\text{IPD}$ & 0.497 & 0.722 & 0.875 & 0.670 & 0.266 & 0.362 & 0.470 \\
  &            & $\text{tPoS}_\text{AGD}$ & 0.536 & 0.752 & 0.891 & 0.691 & 0.233 & 0.339 & 0.431 \\
  &            & $\text{tPoS}_\text{AGD-Q}$ & 0.519 & 0.747 & 0.885 & 0.684 & 0.242 & 0.344 & 0.449 \\
    \\
\multirow{5}{*}{10} & \multirow{5}{*}{\shortstack{Mean shift from\\Phase II to Phase III}} & NoShift & 0.381 & 0.629 & 0.816 & 0.590 & 0.201 & 0.300 & 0.400 \\
 &           & Standard & 0.634 & 0.835 & 0.940 & 0.760 & 0.058 & 0.110 & 0.174 \\
 &           & $\text{tPoS}_\text{IPD}$ & 0.388 & 0.636 & 0.830 & 0.598 & 0.191 & 0.294 & 0.391 \\
 &           & $\text{tPoS}_\text{AGD}$ & 0.362 & 0.606 & 0.818 & 0.577 & 0.212 & 0.318 & 0.425 \\
 &           & $\text{tPoS}_\text{AGD-Q}$ & 0.362 & 0.600 & 0.809 & 0.575 & 0.217 & 0.321 & 0.425 \\
   \\
\multirow{5}{*}{11} & \multirow{5}{*}{\shortstack{Mean and variance shift\\ from Phase II to Phase III}} & NoShift & 0.495 & 0.733 & 0.883 & 0.673 & 0.200 & 0.302 & 0.400 \\
 &           & Standard & 0.285 & 0.524 & 0.739 & 0.513 & 0.413 & 0.537 & 0.648 \\
 &           & $\text{tPoS}_\text{IPD}$ & 0.403 & 0.655 & 0.847 & 0.613 & 0.279 & 0.396 & 0.492 \\
 &           & $\text{tPoS}_\text{AGD}$ & 0.424 & 0.660 & 0.848 & 0.620 & 0.265 & 0.388 & 0.487 \\
 &           & $\text{tPoS}_\text{AGD-Q}$ & 0.411 & 0.657 & 0.846 & 0.615 & 0.282 & 0.395 & 0.486 \\
   \\
\multirow{5}{*}{12} & \multirow{5}{*}{\shortstack{Mean and variance shift\\ from Phase II to Phase III}}          & NoShift & 0.351 & 0.597 & 0.796 & 0.565 & 0.200 & 0.301 & 0.401 \\
 &           & Standard & 0.551 & 0.775 & 0.908 & 0.711 & 0.077 & 0.133 & 0.202 \\
 &           & $\text{tPoS}_\text{IPD}$ & 0.341 & 0.569 & 0.787 & 0.557 & 0.215 & 0.306 & 0.415 \\
 &           & $\text{tPoS}_\text{AGD}$ & 0.321 & 0.540 & 0.769 & 0.540 & 0.235 & 0.333 & 0.441 \\
 &           & $\text{tPoS}_\text{AGD-Q}$ & 0.314 & 0.552 & 0.764 & 0.539 & 0.237 & 0.331 & 0.441 \\
   \\
\multirow{5}{*}{13} & \multirow{5}{*}{\shortstack{Distributional shape shift\\from Phase II to Phase III}} & NoShift & 0.565 & 0.774 & 0.912 & 0.713 & 0.200 & 0.301 & 0.400 \\
 &           & Standard & 0.366 & 0.606 & 0.816 & 0.580 & 0.383 & 0.503 & 0.609 \\
 &           & $\text{tPoS}_\text{IPD}$ & 0.461 & 0.695 & 0.867 & 0.645 & 0.298 & 0.401 & 0.510 \\
 &           & $\text{tPoS}_\text{AGD}$ & 0.362 & 0.604 & 0.806 & 0.577 & 0.400 & 0.504 & 0.608 \\
 &           & $\text{tPoS}_\text{AGD-Q}$ & 0.476 & 0.689 & 0.865 & 0.649 & 0.277 & 0.398 & 0.512 \\
   \\
\multirow{5}{*}{14} &    \multirow{5}{*}{\shortstack{Distributional shape shift\\from Phase II to Phase III}}       & NoShift & 0.373 & 0.615 & 0.809 & 0.583 & 0.200 & 0.300 & 0.401 \\
 &           & Standard & 0.612 & 0.814 & 0.928 & 0.743 & 0.067 & 0.115 & 0.181 \\
 &           & $\text{tPoS}_\text{IPD}$ & 0.373 & 0.617 & 0.815 & 0.586 & 0.192 & 0.293 & 0.390 \\
 &           & $\text{tPoS}_\text{AGD}$ & 0.497 & 0.730 & 0.878 & 0.670 & 0.118 & 0.187 & 0.270 \\
 &           & $\text{tPoS}_\text{AGD-Q}$ & 0.335 & 0.572 & 0.786 & 0.554 & 0.234 & 0.337 & 0.446 \\

    \bottomrule
    \end{tabular}
\end{table}

\begin{table}[!htbp]
\caption{Baseline characteristics for the MUSE and TULIP-1 trials.}
\label{tab:muse_tulip1}
\small
\centering
\begin{tabular}{lcc}
\toprule

\textbf{Characteristic}
& \textbf{MUSE}
& \textbf{TULIP-1} \\

\midrule

Age, years                                   & 39.2 (12.4) & 41.5 (12.2) \\
Female                                       & 92.5\%      & 92.3\%      \\
White race                                   & 37.8\%      & 72.0\%      \\
High type I IFN gene signature                & 75.1\%      & 82.1\%      \\
SLEDAI-2K global score                        & 10.9 (4.1)  & 11.4 (3.8)  \\
Physician's global assessment                & 1.81 (0.42) & 1.85 (0.40) \\
CLASI activity score                         & 7.1 (5.7)   & 8.3 (7.0)   \\
Oral corticosteroid use                      & 83.1\%      & 83.2\%      \\
Oral corticosteroid dose $\geq 10$ mg/day     & 59.2\%      & 56.3\%      \\
Antimalarial use                             & 75.1\%      & 70.9\%      \\

\bottomrule
\end{tabular}

\vspace{2mm}
\begin{flushleft}
\small
Baseline summaries were obtained by pooling the published arm-specific summaries for the anifrolumab 300-mg and placebo groups within each trial according to their sample sizes. Continuous characteristics are presented as mean (SD), and binary characteristics as percentages. 
IFN = interferon; SLEDAI-2K = Systemic Lupus Erythematosus Disease Activity Index 2000; CLASI = Cutaneous Lupus Erythematosus Disease Area and Severity Index.

\end{flushleft}

\end{table}


\clearpage

\end{document}